\UseRawInputEncoding
\documentclass[preprint,nofootinbib,aps,superscriptaddress,eqsecnum]{revtex4-1} 
 \pdfoutput=1
 \usepackage{graphicx}
 \usepackage{amsmath}
\usepackage{amsfonts}
\usepackage{amssymb}
\usepackage{hyperref}
\usepackage{cleveref}
\usepackage{caption}
\usepackage{subcaption}
\def\bea{\begin{eqnarray}}
\def\eea{\end{eqnarray}}
 \def\be{\begin{equation}}
\def\ee{\end{equation}}

\begin{document}

\title{Low Energy Constraints From Absolute Neutrino Mass Observables and Lepton Flavor Violation in Left-Right Symmetric Model}

\author{Srubabati Goswami}
\email[Email Address: ]{sruba@prl.res.in}
\affiliation{Theoretical Physics Division, 
Physical Research Laboratory, Ahmedabad - 380009, India}

 \author{Vishnudath K. N.}
\email[Email Address: ]{vishnudath@prl.res.in}
\affiliation{Theoretical Physics Division, 
Physical Research Laboratory, Ahmedabad - 380009, India}

\begin{abstract}
We have studied the correlations among the three absolute neutrino mass observables - the effective Majorana mass ($m_{ee}$) which can be obtained from neutrinoless double beta decay, the electron neutrino mass ($m_{\beta}$) which is measured in single beta decay experiments and the sum of the light neutrino masses ($\Sigma$) which is constrained from cosmological observations, in the context of minimal left-right symmetric model. Two phenomenologically interesting  cases of type-I seesaw dominance as well as type-II seesaw dominance have been considered. We have taken into account the independent constraints coming from lepton flavor violation, single $\beta$ decay, cosmology and neutrinoless double beta decay and have determined the combined allowed parameter space that can be probed in the future experiments. We have also analyzed the correlations and tensions between the different mass variables. In addition, the constraints on the masses of  the heavy particles coming from lepton flavor violation and the bounds on three absolute neutrino mass observables are also determined. We show that these constraints can rule out some of the parameter space which are not probed by the collider experiments.

\end{abstract}
 
\maketitle

\section{Introduction}

The Standard Model (SM) of particle physics, despite being a highly successful theory in many aspects,  has certain limitations, which motivates one to think of scenarios beyond SM. One of the main limitations is the absence of the neutrino mass in SM. The observation of neutrino oscillations has shown that neutrinos can convert from one flavour to another implying non-zero masses. The global analysis of the neutrino oscillation data from various solar, atmospheric, reactor and accelerator experiments has determined the values of two mass squared differences ($\Delta m_{sol}^2 \sim 7.4 \times 10^{-5} ~eV^2$ and $\Delta m_{atm}^2 \sim 2.5 \times 10^{-3}~eV^2$), and three neutrino mixing angles ($\theta_{12} \sim 34^{\circ} $, $\theta_{23} \sim 48 ^{\circ}$ and $\theta_{13} \sim 7.4^{\circ}$ )\cite{Capozzi:2017ipn,deSalas:2017kay,Esteban:2018azc}.

In addition, we also do not know if neutrinos are Dirac particles or lepton number violating Majorana particles. This can be probed in the neutrinoless double beta decay ($0\nu\beta\beta$) experiments, which if observed, would prove that neutrinos are Majorana particles \cite{Furry:1939qr,Schechter:1980gr}. Also, the $0\nu\beta\beta$ experiments are sensitive to the absolute neutrino masses through the dependence on the effective Majorana mass $m_{ee}$, unlike the oscillation experiments which tell us only about the mass squared differences. The combined constraints from KamLAND-Zen and GERDA experiments put an upper limit on $m_{ee}$ in the range $0.071-0.161$ eV depending on the values of the Nuclear Matrix Elements (NMEs) \cite{Agostini:2018tnm}. Other than $0\nu\beta\beta$, the non-oscillation data from single $\beta$ decay and the cosmological observations put independent constraints on the absolute neutrino masses. The rate of single $\beta$ decay depends on the electron neutrino mass $m_\beta$ and this is constrained to be less than 1.1 eV by the KATRIN experiment~\cite{Aker:2019uuj} and the Planck-2018 data constrains the sum of active light neutrino masses, $\Sigma$ to be less that 0.12 eV~\cite{Aghanim:2018eyx}. It is important to study the combined constraints coming from these mutually exclusive experiments and make predictions that can be tested in the future experiments~\cite{Fogli:2004as,Pascoli:2005zb,Fogli:2008ig,DiIura:2016zsx,
Xing:2016ymd,Caldwell:2017mqu,Agostini:2017jim}.

One of the most popular and elegant way of generating neutrino masses is the seesaw mechanism. According to this, neutrinos are Majorana particles whose masses are generated by the dimension-5 Weinberg operator $\kappa l_L l_L \Phi \Phi $ \cite{weinberg} through electroweak symmetry breaking. Depending on the way in which the Weinberg operator is generated, there are three types of seesaw mechanisms : (i) type-I seesaw mediated by heavy right-handed neutrinos \cite{Minkowski:1977sc,seesaw1,seesaw2,Mohapatra:1979ia}, (ii) type-II seesaw mediated by heavy scalar triplet \cite{Schechter:1980gr,Schechter:1981cv,Lazarides:1980nt,
Mohapatra:1980yp} and (iii) type-III seesaw mediated by heavy fermionic triplets \cite{Foot:1988aq}. There are many gauge extensions of the SM that incorporate seesaw mechanism. Among these, the left-right symmetric models (LRSMs) based on the gauge group $SU(3)_C \times SU(2)_R \times SU(2)_L \times U(1)_{B-L}$ are of special significance \cite{Pati:1974yy,Mohapatra:1974gc,Senjanovic:1975rk,
Mohapatra:1980qe}. Here, the origin of parity violation in SM is understood in a simple way such that the parity symmetry is restored at a higher energy scale. Also, these models contain right-handed neutrinos naturally and hence can generate neutrino masses by type-I seesaw mechanism. Such models can be embedded in grand unified theories based on the gauge group $SO(10)$ and the presence of the $SU(2)_R$ gauge bosons and additional scalars makes their study interesting from an experimental point of view. There are two most popular variants of the LRSM which differ from each other in their Higgs sector. One of them contains two Higgs doublets in addition to the Higgs bidoublet needed for the SM symmetry breaking whereas the other variant has Higgs triplets instead of the Higgs doublets. The model with Higgs triplets can also contribute to neutrino masses via type-II seesaw mechanism. In addition, the associated charged scalars can have interesting signatures in the collider experiments \cite{Gunion:1995mq,Huitu:1996su,Huitu:1997vh,Dev:2016dja,Aaboud:2017qph,Borah:2018yxd}. In this paper, we consider the LRSM with triplet scalars and from here onwards, we refer to this model as the minimal left-right symmetric model (MLRSM).

Another important feature of the  MLRSM is that there can be a number of new physics contributions to  $0\nu\beta\beta$, coming from right-handed neutrinos as well as the Higgs triplets, especially when these particles are at the TeV scale \cite{Hirsch:1996qw,Tello:2010am,Chakrabortty:2012mh,Barry:2013xxa,
Dev:2014xea,Awasthi:2015ota,
Bambhaniya:2015ipg,Bonilla:2016fqd,Awasthi:2016kbk,Deppisch:2017vne,
Borah:2017ldt,Borgohain:2017akh}. These diagrams correspond to the short-range mechanism, unlike the long range mechanism due to the standard light neutrino exchange. In such cases, there can be an enhancement or suppression of the rate of $0\nu\beta\beta$ due to the interference of different diagrams. One can study various limiting cases where the new physics contributions can be expressed in terms of a few free parameters along with the light neutrino masses and the Pontecorvo-Maki-Nakagawa-Sakata (PMNS) mixing matrix \cite{Tello:2010am,Chakrabortty:2012mh,Bambhaniya:2015ipg}.
The implications of TeV scale MLRSM for $0\nu\beta\beta$ under various limits and approximations have been studied in \cite{Tello:2010am,Chakrabortty:2012mh,Barry:2013xxa,Dev:2013vxa,
Dev:2014xea,Awasthi:2015ota,
Bambhaniya:2015ipg,Awasthi:2016kbk,Bonilla:2016fqd,Deppisch:2017vne,
Borah:2017ldt,Borgohain:2017akh}. In particular, the authors of \cite{Bambhaniya:2015ipg,Bonilla:2016fqd} have studied the combined constraints from $0\nu\beta\beta$ and lepton flavour violation (LFV) in the context of MLRSM by taking into account the scalar triplet contributions. It has been observed that the LFV bounds put constraints on the absolute neutrino masses and, it will be interesting to see how these constraints are reflected in the allowed regions in the planes of the three mass variables, $m_\beta,~ m_{ee} $ and $\Sigma$.

In this paper, we study the correlations among the three observables, $m_{ee}$, $m_{\beta}$ and $\Sigma$ which depend on the absolute neutrino masses, in MLRSM, for both type-I as well as type-II seesaw dominant cases.
Even though many such studies have been done in the case of standard three neutrino picture~\cite{Fogli:2004as,Pascoli:2005zb,Fogli:2008ig,DiIura:2016zsx,
Xing:2016ymd,Caldwell:2017mqu,Agostini:2017jim}, no detailed analysis has been done for MLRSM.
 In particular, we have taken the independent constraints coming from  LFV, cosmology and $0\nu\beta\beta$ decay into account and obtained the allowed parameter spaces that can be probed in the future experiments. In addition, we study how the LFV constraints and the bounds on  $m_{ee}$, $m_{\beta}$ and $\Sigma$ restrict the masses of  the heavy neutrinos, the charged triplet scalar as well as the $W_R$ gauge boson. In fact, we find that the LFV constraints along with the bounds on absolute neutrino mass observables can rule out some of the parameter space which are not probed by the collider experiments, thus providing a complimentary way to study the new physics parameters.

The rest of the paper is organized as follows. In section-II, we describe the  MLRSM, it's particle content and the mechanism of neutrino mass generation. The rates of the LFV decays in this model and the constraints coming from them are discussed in section-III. In section-IV, the predictions for $0\nu\beta\beta$  in MLRSM are discussed and in section-V, we have discussed the correlations among  $m_{ee}$, $m_{\beta}$ and $\Sigma$. Constrains on the new physics parameters from LFV and neutrino mass observables are discussed in section-VI. Finally, we summarize our findings in section-VII.

\section{Minimal left-right symmetric model}

The  MLRSM is based on the gauge group $SU(3)_C \times SU(2)_R \times SU(2)_L \times U(1)_{B-L}$. The matter multiplets for quarks and leptons in this model are given by,
\be Q_L \,\, \sim  \,\,(2, 1, \frac{1}{3}) \,\,\,\, ; \,\,Q_R \,\,\sim \,\, (1, 2, \frac{1}{3}), \nonumber \ee
\be l_L \,\, \sim \,\, (2, 1, -1) \,\,\,\, ; \,\,l_R \,\, \sim \,\, (1, 2, -1) . \nonumber \ee
Thus, the existence of right handed neutrinos as part of the $SU(2)_R$ doublet $l_R$ implies that the active light neutrino masses can be generated via a type-I seesaw mechanism. The Higgs Sector consists of,
 \be \Phi  \,\, \sim \,\, (2,2,0) \,\,\,\, ;\,\, \,\, \Delta_R \,\,\sim\,\,(1,3,2) \,\, \,\,; \,\,\,\, \Delta_L \,\, \sim \,\,(3,1,2).\nonumber \ee
 
 Here, $\Delta_R$ is the $SU(2)_R$ triplet that breaks the $SU(2)_R \times SU(2)_L \times U(1)_{B-L}$ gauge group to the SM gauge group $ SU(2)_L \times U(1)_{Y}$. The bidoublet $\Phi$ is responsible for breaking the SM symmetry group to $U(1)_{em}$. The presence of the $SU(2)_L$ triplet Higgs is required by left-right symmetry and it gives rise to type-II seesaw mechanism. Thus, in general, MLRSM can have type- I+II seesaw mechanism for neutrino mass generation. The Higgs fields acquire vevs as,
\be \langle \Phi \rangle \,=\,\begin{pmatrix}
 v & 0 \\
 0 & v'
\end{pmatrix} \,\,\, ; \,\,\,\langle \Delta_L \rangle \,=\, v_L \,\,\, ;\,\,\, \langle \Delta_R \rangle \,=\, v_R .\ee
The part of the Lagrangian that is relevant for the generation of neutrino mass is,
\be   L_l = f_L\,\bar{l}_L^c\,\Delta_L\,l_L \,+\,  f_R\,\bar{l}_R^c\,\Delta_R\,l_R \,\,+\, \bar{l}_R(Y_D \Phi \,+\,Y_L\tilde{\Phi})l_L \,\, + \textrm{h.c.}   .  \ee
Note that here, the generation indices have been suppressed and that $Y_D $, $f_L$ and $f_R$ are $3 \times 3$ matrices. Depending upon whether we take the symmetry between the left and right sectors to be parity or charge conjugation, the Yukawa coupling matrices for the left- and the right-handed leptons with the corresponding scalar triplets will be related as,
\be f_L = f_R \,\,\, \textrm{or} \,\,\, f_L = f_R^*, \ee respectively~\cite{Maiezza:2010ic}. In our work, we take parity as the symmetry between left and right sectors. Once the Higgs fields acquire vevs, we will get the neutrino mass matrix in the basis $(\nu_L^c\,,\, N_R)$  as,
\be \begin{pmatrix}
 f_Lv_L & Y_Dv \\
 Y_D^T v & f_R v_R
\end{pmatrix} .\ee
The neutrino mass is generated through seesaw mechanism $(f_Rv_R >>Y_D v)$ and we get light and heavy neutrino mass matrices as,
\be m_\nu  =  f_Lv_L - \frac{v^2}{v_R} Y_D^T\,f_R^{-1}Y_D\ee and 
\be \,\,M_R \,\, =\,\, f_Rv_R \label{heavy}\ee
respectively. Now we can consider two interesting limits~\cite{Tello:2010am,Chakrabortty:2012mh,Bambhaniya:2015ipg} :

\paragraph{\bf Type-I Dominance} : In this case, $f_Lv_L << \frac{v^2}{v_R} Y_D^T\,f_R^{-1}Y_D$ and hence, the light neutrino mass matrix is given as,
\be m_\nu  =  - \frac{v^2}{v_R} Y_D^T\,f_R^{-1}Y_D.\ee
 Assuming $Y_D \propto I$, where $I$ is the $3 \times 3$ identity matrix, one can see that the unitary matrices diagonalizing the light and the heavy neutrino mass matrices are related as,
 \be V = U^*, \label{Urelation} \ee
where $V$ is the unitary matrix diagonalizing the heavy neutrino mass matrix $M_R$ and $U$ is the UPMNS matrix that diagonalizes $m_\nu$. Also, the heavy neutrino masses are inversely proportional to the light neutrino masses, i.e.,
 \be M_{N_i} \propto \frac{1}{m_i} .\ee

\paragraph{\bf Type-II Dominance} : In this case, $f_Lv_L >> \frac{v^2}{v_R} Y_D^T\,f_R^{-1}Y_D$ and hence, the light neutrino mass matrix is given as,
\be m_\nu  =  f_Lv_L = f_Rv_L,\ee
where the last equality follows from left-right symmetry. Thus, the light and the heavy neutrino mass matrices are diagonalized by the same unitary matrix, i.e.,
\be V = U .\ee
But now, the heavy neutrino masses are directly proportional to the light neutrino masses, i.e.,
 \be M_{N_i} \propto m_i .\ee
 In our work, we denote the mass of the heaviest right handed neutrino as $M_N$.

 Another assumption that we will be making is in the scalar sector. In scalar sector, there are 20 real degrees of freedom (8 from the bidoublet, 6 from $\Delta_L$ and 6 from $\Delta_R$), out of which, 6 will become Goldstone bosons and one will be the physical SM Higgs particle with mas $\sim 125$ GeV. The remaining physical scalars are the doublets $H_\phi^0, A_\phi^0, H_\phi^\pm$, the left triplets $H_L^0, A_L^0 ,\Delta_L^\pm, \Delta_L^{\pm \pm}$ and the right triplets $H_R^0, A_R^0, \Delta_R^\pm, \Delta_R^{\pm \pm}$. All of these particles are heavy and we assume that the masses of the doubly charged scalars to be the same in left and right sectors. We define the parameter $M_\Delta$ in terms of the masses of the two doubly charged scalars,
 \be \frac{1}{M_\Delta^2} = \frac{1}{{M_{\Delta_L^{\pm\pm}} }^2} + \frac{1}{{M_{\Delta_R^{\pm\pm}} }^2} , \label{mscalar}\ee
 and take $r  =  M_N/M_\Delta$, where $r$ is a fixed proportionality constant. Thus we have,
\be  M_{\Delta_R^{++}} = \sqrt{2} M_N/r .\ee

\section{Constraints from Lepton Flavour Violation}

The presence of the $SU(2)_R$ gauge bosons, triplet scalars and right handed neutrinos can give rise to observable rates of charged LFV decays that can be tested in future experiments. Ignoring the light-heavy mixing in the neutrino sector as well as the left-right gauge boson mixing, the branching ratio for the LFV decay $\mu \rightarrow e \gamma$ in MLRSM is given as \cite{Cirigliano:2004mv,Barry:2013xxa},
\be BR (\mu \rightarrow e \gamma) =  \frac{3 \alpha_{em}}{2 \pi} \Big( |G_L^\gamma|^2 + |G_R^\gamma|^2 \Big) .\ee
In the above equation,
\be G_R^\gamma = V_{\mu i} V_{e i}^*~ \Big(~ \frac{M_{W_L}^2}{M_{W_R}^2} G_1^\gamma(b_i) + \frac{2b_i}{3} \frac{M_{W_L}^2}{{M_{\Delta_R^{++}} }^2} ~ \Big)\ee
and,
\be G_L^\gamma = V_{\mu i} V_{e i}^* ~b_i \Big(~  \frac{2}{3} \frac{M_{W_L}^2}{{M_{\Delta_L^{++}} }^2} + \frac{1}{12} \frac{M_{W_L}^2}{{M_{\Delta_L^{+}} }^2} ~ \Big),\ee
where $a_i = \Big( \frac{M_{N_i}}{M_{W_L}}  \Big)^2$, $b_i = \Big( \frac{M_{N_i}}{M_{W_R}}  \Big)^2$, $\alpha_{em} \approx 1/137$,  $M_{W_L}$ and $M_{W_R}$ are the masses of left- and right- gauge bosons respectively, $V$ is the unitary matrix that diagonalizes the right-handed neutrino mass matrix, $M_{N_i}$ are the masses of heavy right-handed neutrinos, and,
\be G_1^\gamma (a) = \frac{-2a^3 + 5a^2 -a}{4(1-a)^3} - \frac{3a^3}{2(1-a)^4}~\textrm{ln}a .\ee
The current upper bound on  the branching ratio for $(\mu \rightarrow e \gamma)$ is \cite{TheMEG:2016wtm}, 
\be BR(\mu \, \rightarrow e \, \gamma)  \,\, < \, 4.2 \times 10^{-13} . \label{brmeg}\ee
The upper bound on the branching ratio for the decay $\mu \rightarrow 3 e $ also puts very strong constraints on the masses of the heavy particles in low scale seesaw models as\cite{Bertl:1985mw},
\be \textrm{Br}(\mu \, \rightarrow 3 e)  \,\, < \,1.0 \times 10^{-12} .\ee
The expression for the branching ratio of $\mu \rightarrow 3 e$ in MLRSM is given as \cite{Cirigliano:2004mv,Leontaris:1985qc,
Swartz:1989qz,Cirigliano:2004tc},
\be  BR(\mu \rightarrow 3 e) = \frac{1}{2} |h_{\mu e} h_{ee}^*|^2 \Big(    \frac{M_{W_L}^4}{{M_{\Delta_L^{++}} }^4} +  \frac{M_{W_L}^4}{{M_{\Delta_R^{++}} }^4}  \Big), \ee
where,
\be h_{\alpha \beta } = \sum_{i=1}^{3} V_{\alpha i} V_{\beta i} \frac{M_{N_i}}{M_{W_R}} . \ee


\section{$0\nu\beta\beta$ in  MLRSM}

There are several extra contributions to $0\nu\beta\beta$ in MLRSM, coming from right-handed currents, left-right mixing and scalar triplets. All these will give extra contributions to the effective Majorana mass which can be directly probed in $0\nu\beta\beta$ experiments. Assuming that there are no large light-heavy neutrino as well as left-right gauge boson mixings, the half life for $0\nu\beta\beta$ is given as~\cite{Tello:2010am},
\be \frac{1}{T_{1/2}^{0\nu}} = G_{01}^{0\nu} \Big( |M_\nu^{0\nu} \eta_\nu|^2 +  |M_N^{0\nu} \eta_R|^2  \Big). \ee
Here, $M_\nu^{0\nu}$ and $M_N^{0\nu}$ are the nuclear matrix elements for the light- and heavy-neutrino exchanges respectively and $G_{01}^{0\nu}$ is the phase space factor. $\eta_\nu$ and $\eta_R$ represent the left-handed and right-handed amplitudes respectively, and are given as,
\be \eta_\nu = \frac{1}{m_e}\sum_i U_{ei}^2 m_i \,\,\,\,;\,\,\,\, \eta_R = m_p \Big( \frac{M_{W_L}}{M_{W_R}} \Big)^4 \Big(\sum_i \frac{V^2_{ei} }{M_{N_i}} +  \sum_i \frac{V^2_{ei} M_{N_i} }{M_{\Delta_R^{++}}^2}\Big)  .\ee Here, $m_e$ is the electron mass and $m_p$ is the proton mass. The corresponding effective Majorana neutrino mass is given as,
\be  m_{ee} = \sqrt{\Big|\sum_i U_{ei}^2 m_i \Big|^2 + \Big|\langle p^2 \rangle \Big( \frac{M_{W_L}}{M_{W_R}} \Big)^4 \Big(\sum_i \frac{V^2_{ei} }{M_{N_i}} +  \sum_i \frac{V^2_{ei} M_{N_i} }{M_{\Delta_R^{++}}^2}\Big)\Big|^2 }.\label{mee}\ee
Here, $\,\, \langle p^2 \rangle =  |m_e m_p M_N^{0\nu}/M_\nu^{0\nu}| $ encapsulates the contribution due to the nuclear matrix element. Its value is $\sim (153-184 ~\textrm{MeV})^2$ for ${}^{76}Ge$ isotope~\cite{Meroni:2012qf}. Note that we have not included the contribution from the $W_L - W_R$ mixing diagram in the above expression. Recently, in reference~\cite{Li:2020flq}, the authors have emphasized the importance of long-range pion exchange diagrams. Then, the contribution from $W_L-W_R$ mixing can be significant and can dominate over all the other contributions to $0\nu\beta\beta$. In particular, the authors of reference~\cite{Li:2020flq} considered charge conjugation as the left-right symmetry in which case, the $W_L-W_R$ mixing (which is proportional to $\textrm{sin}2\beta$ where $\textrm{tan}\beta = v'/v$) can be large. In our case, we consider parity as the left-right symmetry for which $\textrm{sin}2\beta$ is restricted~\cite{Senjanovic:2014pva,Senjanovic:2015yea} and the $W_L-W_R$ mixing diagram can be neglected when $\beta$ is very small.

\begin{figure}[tbh]
\begin{tabular}{cc}
\includegraphics[width=0.5\textwidth]{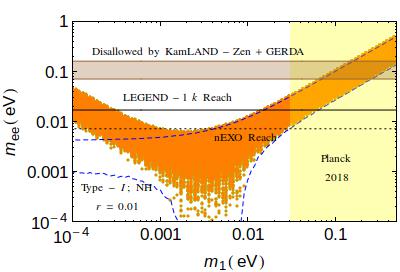}
\includegraphics[width=0.5\textwidth]{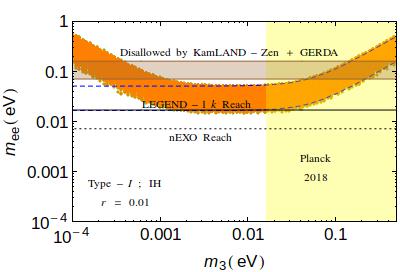}\\
\includegraphics[width=0.5\textwidth]{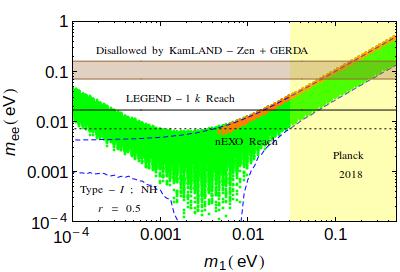}
\includegraphics[width=0.5\textwidth]{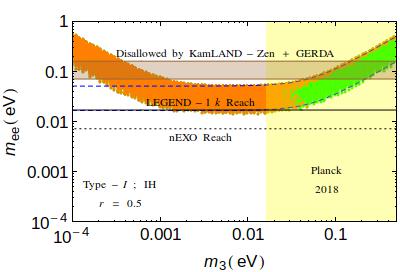}
\end{tabular}
\caption{Predictions for $m_{ee}$ in  MLRSM with type-I dominance for $M_N = 500$ GeV. The left (right) panel is for NH (IH) and the upper (lower) panel is for $r=0.01$ ($r=0.5$). We have fixed $M_{W_R} = 7$ TeV. The Green region is without including the LFV constraints and the orange region satisfies the LFV constraints. The blue dashed lines correspond to the predictions in SM with three light Majorana neutrinos. The yellow region is disfavored by Planck-2018~\cite{Aghanim:2018eyx} and the region above the brown band is disfavored by the combined constraints from KamLAND-Zen and GERDA experiments~\cite{Agostini:2018tnm}. The horizontal dotted black line corresponds to the future sensitivity of nEXO~\cite{Kharusi:2018eqi} and the solid black line corresponds to the future sensitivity of LEGEND-1k~\cite{Abgrall:2017syy,Agostini:2017jim}.}\label{fig1}
\end{figure}
  \begin{figure}[tbh]
\begin{tabular}{cc}
\includegraphics[width=0.5\textwidth]{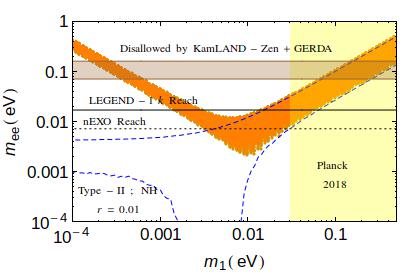}
\includegraphics[width=0.5\textwidth]{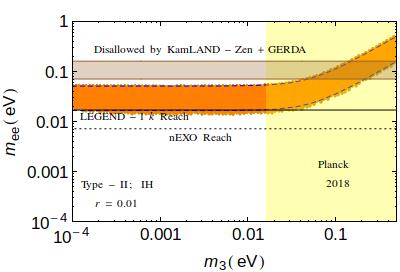}\\
\includegraphics[width=0.5\textwidth]{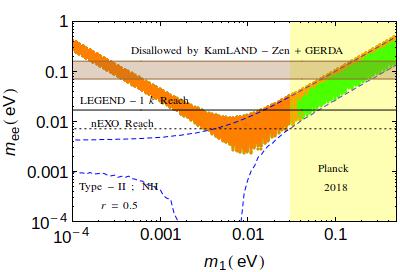}
\includegraphics[width=0.5\textwidth]{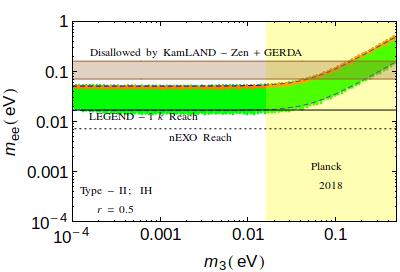}
\end{tabular}
\caption{Same as Fig.\ref{fig1}, but for type-II dominance.}\label{fig2}
\end{figure}

In Fig.\ref{fig1}, we have shown the predictions for $m_{ee}$ in  MLRSM with type-I dominance. The left and the right panels are for normal hierarchy (NH) and inverted hierarchy (IH) of the light neutrino masses respectively, and the upper and the lower panels are for $r=0.01$ and $r=0.5$ respectively. In all the plots in this paper, we have kept $M_{W_R} = 7$ TeV and $M_N = 500$ GeV unless otherwise mentioned. This implies that the upper and the lower panels are for $M_{\Delta_R^{++}} \sim 70 $ TeV and $M_{\Delta_R^{++}} \sim 1.4 $ TeV respectively. The ATLAS search for same-
sign dileptonic new physics signals has set lower bounds on the masses of $\Delta_L^{++}$ and $\Delta_R^{++}$\cite{Aaboud:2017qph}. These are given as $M_{\Delta_L^{++}} > 770-870$ GeV and $M_{\Delta_R^{++}} > 660-760$ GeV, assuming $BR(\Delta^{\pm\pm} \rightarrow l^\pm l^\pm ) = 100 \%$. These chosen values are also allowed by the constraints from perturbativity as well as collider experimens as given in reference \cite{Maiezza:2016ybz, Chauhan:2018uuy}. We have varied the light neutrino mass squared differences and the mixing angles in their $3\sigma$ ranges \cite{Esteban:2018azc} and the Majorana phases have been varied in the range $0-\pi$ for all the plots in this paper. The yellow region is disfavored by Planck-2018 data ($\Sigma = m_1 + m_2 + m_3 < 0.12$ eV)~\cite{Aghanim:2018eyx} and the region above the brown band ($0.071-0.161$ eV) is disfavored by the combined constraints from KamLAND-Zen and GERDA experiments \cite{Agostini:2018tnm}. This is a band because of the NME uncertainty \cite{Engel:2016xgb,Agostini:2018tnm,Kotila:2012zza}.

The green regions in Fig.\ref{fig1} are the predictions for the MLRSM with out including LFV bounds whereas the orange regions satisfy the LFV bounds. The blue dashed lines give the predictions for the SM case with three light Majorana neutrinos. First let us consider the case of NH with type-I dominance (left panels). In the upper panel, the orange and the green regions almost overlap implying that there are no additional constraints from LFV. However, for $r=0.5$, values of $m_1 < 0.0045$ eV are disfavored by LFV constraints and only a narrow region above is allowed by both LFV and $0\nu\beta\beta$ since the Majorana phases corresponding to the region below the orange band give large values of LFV decay rates~\cite{Bambhaniya:2015ipg}. A part of the parameter space is also disfavored by the Planck data. We can see that the cancellation region for NH in MLRSM with type-I dominance gets shifted slightly to the low $m_1$ region as compared to the standard three neutrino scenario. Also, for large values of $m_1$ (the quasi-degenerate region), the predictions for $m_{ee}$ in both the cases are same whereas for smaller values of $m_1$, MLRSM with type-I dominance gives higher values as compared to the three neutrino picture~\cite{Chakrabortty:2012mh,Bambhaniya:2015ipg}.

 Now let us consider IH with type-I dominance (right panel of Fig.\ref{fig1}). Here also, the orange region completely overlaps with the green region for $r=0.01$ and hence all the points are allowed by LFV. But for $r=0.5$, only a narrow region above is allowed by LFV for $m_3 > 0.04 $ eV whereas all the points with  $m_3 < 0.04 $ eV are allowed. The region $m_3 > 0.04 $ eV is also disfavored by Planck-2018 data. We can also see that there is no cancellation region as in the standard three neutrino picture~\cite{Chakrabortty:2012mh,Bambhaniya:2015ipg}. In addition, we can see that the predictions for $m_{ee}$ in both the cases (MLRSM and standard three generation) are same for large values of $m_3$ whereas for smaller values of $m_3$, MLRSM gives much higher values than that in the three neutrino picture, a part of which is disfavored by the bound from KamLand-Zen and GERDA. Thus for type-I dominance with larger values of $r$ ($r = 0.5$), lower values of $m_{lightest}$ are constrained by LFV in the case of NH, whereas, higher values of $m_{lightest}$ are disfavored by LFV in the case of IH.

The horizontal dotted black line in Fig.\ref{fig1} corresponds to the future sensitivity of nEXO, which is a $0\nu\beta\beta$ experiment using ${}^{136}Xe$. The future $3\sigma$ sensitivity of the  nEXO is $T_{1/2} = 5.7 \times 10^{27}$ years~\cite{Kharusi:2018eqi}. This can be converted into a band in $m_{ee}$ by including the NME uncertainties whereas we have given the lowest value in this band ($m_{ee} = 0.007$ eV) by using the highest value of NME. Similarly, the horizontal solid black line corresponds to the future sensitivity of LEGEND-1k~\cite{Abgrall:2017syy,Agostini:2017jim}. LEGEND-1k is a $0\nu\beta\beta$ experiment using ${}^{76}Ge$ and has a $3\sigma$ discovery sensitivity of $T_{1/2} = 4.5 \times 10^{27}$ years. This translates into a band in $m_{ee}$ with lowest value as $0.017$ eV. From the figure, we can see that if the LGEND-1k does not observe a signal in the next run, it will rule out IH completely for both the considered values of $r$ (or $M_{\Delta_R^{++}}$). This is true for MLRSM as well as the standard three neutrino picture. Also, LEGEND-1k and nEXO can rule out some part of the low $m_1$ region for MLRSM with $r=0.01$ if no signal is observed.

In Fig.\ref{fig2}, we have shown similar plots as in Fig.\ref{fig1}, but for type-II dominance. The interesting feature to note here is the absence of cancellation region for NH (left panel). Here also, the green and the orange regions overlap completely for $r=0.01$ (which is true for IH as well, as can be seen from the right panel). For $r=0.5$, the values of $m_{ee}$ in the green region with $m_1 > 0.048$ eV are constrained by LFV unlike in the type-I dominance case where lower values of $m_1$ are disfavored for NH. This is also disfavored by Planck 2018~\cite{Aghanim:2018eyx}.  On the other hand, the predictions for IH is almost the same as in the three neutrino picture. This is because we have taken $M_{W_R} = 7$ TeV. For smaller values of $M_{W_R}$, the predictions for type-II IH would have been different from the standard three generation case~\cite{Tello:2010am,Chakrabortty:2012mh,Bambhaniya:2015ipg}. For IH with $r=0.5$, the values of predicted $m_{ee}$ in the green region are disfavored by LFV through the entire range of $m_3$ since the corresponding values of Majorana phases are disfavored. From the left panel of Fig.\ref{fig2}, we can see that a part of the parameter space for low $m_1$ is already disfavored by the current constraints on $0\nu\beta\beta$ for NH in MLRSM. Also, IH with $r=0.5$ will be disallowed if LEGEND-1k fails to observe a positive signal and NH with $r=0.5$ will be be disallowed if nEXO fails to observe a positive signal.

\section{Correlations of $m_{ee}$ to $m_\beta$ and $\Sigma$}
  \begin{figure}[tbh]
\begin{tabular}{cc}
\includegraphics[width=0.5\textwidth]{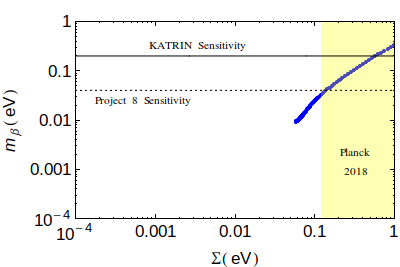}
\includegraphics[width=0.5\textwidth]{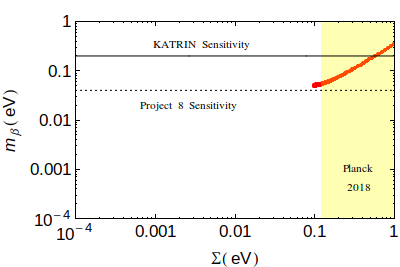}
\end{tabular}
\caption{Correlation of $m_\beta$ against $\Sigma$. The left (right) panel is for NH (IH). The yellow region is disfavored by Planck-2018~\cite{Aghanim:2018eyx}. The black solid and dotted lines correspond to the future sensitivity of KATRIN~\cite{Aker:2019uuj} and Project 8~\cite{Esfahani:2017dmu} respectively.}\label{fig3}
\end{figure}
  \begin{figure}[tbh]
\begin{tabular}{cc}
\includegraphics[width=0.5\textwidth]{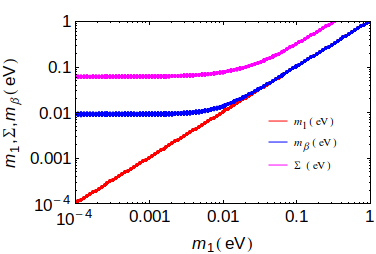}
\includegraphics[width=0.5\textwidth]{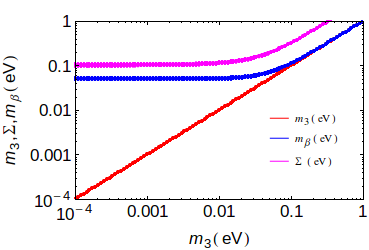}
\end{tabular}
\caption{Lightest neutrino mass (Red), $m_\beta$ (blue) and $\Sigma$ (Magenta) plotted against the lightest neutrino mass. The left (right) panel is for NH (IH).}\label{fig3b}
\end{figure}

  \begin{figure}[tbh]
\begin{tabular}{cc}
\includegraphics[width=0.5\textwidth]{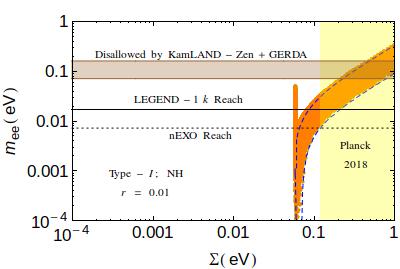}
\includegraphics[width=0.5\textwidth]{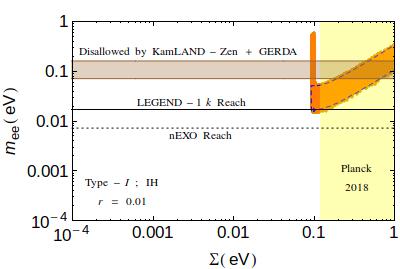}\\
\includegraphics[width=0.5\textwidth]{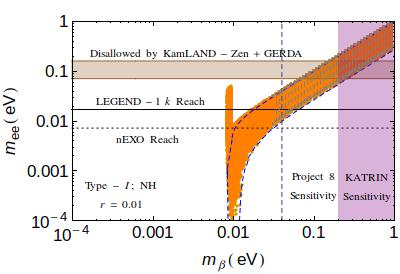}
\includegraphics[width=0.5\textwidth]{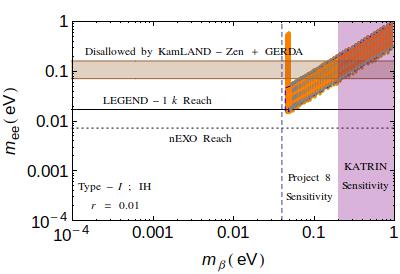}
\end{tabular}
\caption{Correlations of $m_{ee}$ against $m_\beta$ and $\Sigma$ for type-I dominance with $M_N = 500$ GeV, $r=0.01$ ($M_{\Delta_R^{++}} \sim 70 $ TeV) and $M_{W_R} = 7$ TeV. The Green region is without demanding the LFV constraints which is completely covered by the the orange region for which LFV constraints are satisfied. The blue dashed lines correspond to the predictions in standard three neutrino picture. The yellow region in the upper panels and the region shaded with gray dashes in the lower panels are disfavored by Planck-2018 and the region above the brown band is disfavored by the constraints from KamLAND-Zen and GERDA~\cite{Agostini:2018tnm}. The purple region corresponds to the future sensitivity of KATRIN~\cite{Aker:2019uuj} and the vertical dashed black line corresponds to the future sensitivity of Project 8~\cite{Esfahani:2017dmu}. The horizontal dotted and solid black lines correspond to the future sensitivities of nEXO~\cite{Kharusi:2018eqi} and LEGEND-1k~\cite{Abgrall:2017syy,Agostini:2017jim} respectively.}\label{fig4}
\end{figure}
  \begin{figure}[tbh]
\begin{tabular}{cc}
\includegraphics[width=0.5\textwidth]{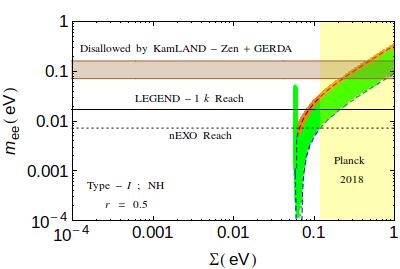}
\includegraphics[width=0.5\textwidth]{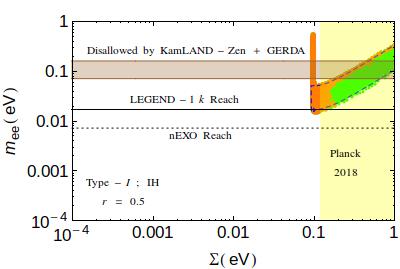}\\
\includegraphics[width=0.5\textwidth]{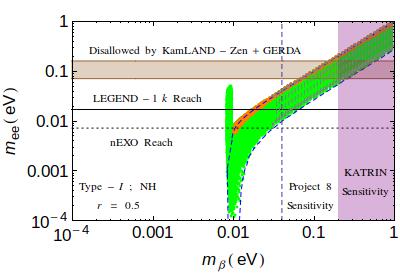}
\includegraphics[width=0.5\textwidth]{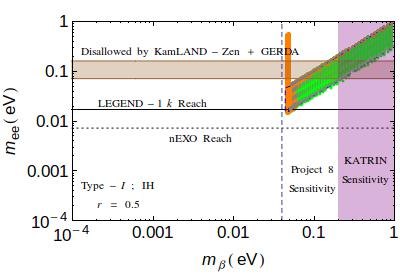}
\end{tabular}
\caption{Same as Fig.\ref{fig4}, but for type-I dominance with $r=0.5$ ($M_{\Delta_R^{++}} \sim 1.4 $ TeV).}\label{fig5}
\end{figure}
  \begin{figure}[tbh]
\begin{tabular}{cc}
\includegraphics[width=0.5\textwidth]{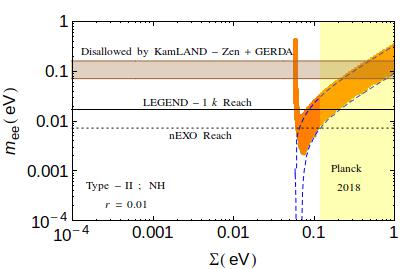}
\includegraphics[width=0.5\textwidth]{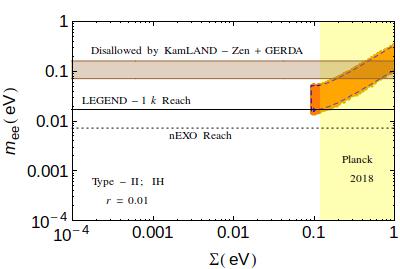}\\
\includegraphics[width=0.5\textwidth]{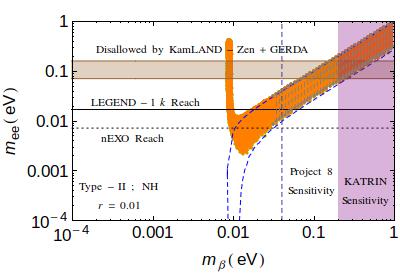}
\includegraphics[width=0.5\textwidth]{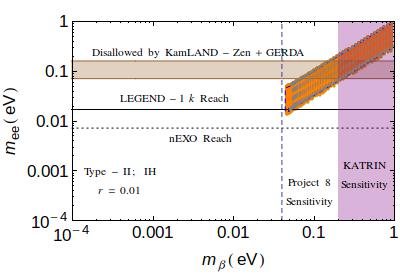}
\end{tabular}
\caption{Same as Fig.\ref{fig4}, but for type-II dominance with $r=0.01$ ($M_{\Delta_R^{++}} \sim 70 $ TeV).}\label{fig6}
\end{figure}
  \begin{figure}
\begin{tabular}{cc}
\includegraphics[width=0.5\textwidth]{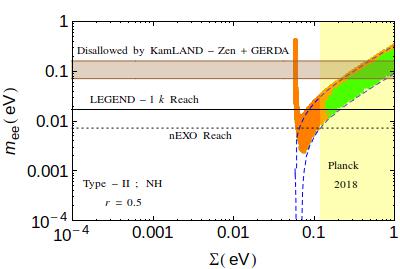}
\includegraphics[width=0.5\textwidth]{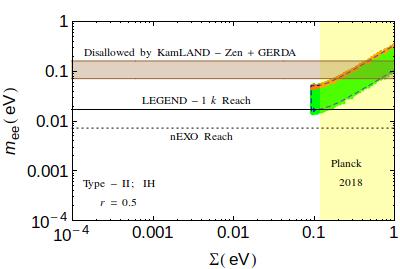}\\
\includegraphics[width=0.5\textwidth]{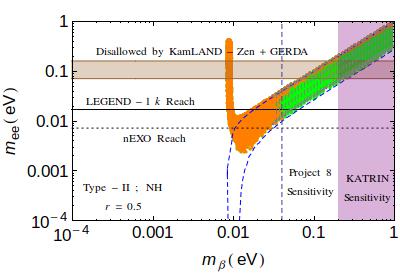}
\includegraphics[width=0.5\textwidth]{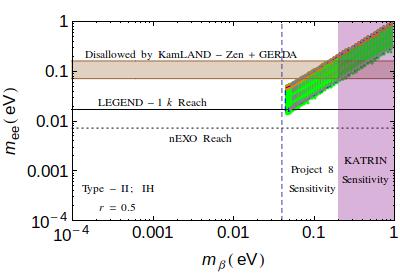}
\end{tabular}
\caption{Same as Fig.\ref{fig4}, but for type-II dominance with $r=0.5$ ($M_{\Delta_R^{++}} \sim 1.4 $ TeV).}\label{fig7}
\end{figure}

As it was already mentioned, in addition to $0\nu\beta \beta$, the non-oscillation data from single $\beta$ decay and cosmology put independent constraints on the absolute neutrino masses. The single $\beta$ decay is sensitive to the electron neutrino mass, $m_\beta$, which is defined as,
\be m_\beta = \sqrt{\sum_i |U_{ei}^2 m_i^2|}. \ee
The most stringent limit on $m_\beta$ has been obtained recently by KATRIN (Karlsruhe Tritium Neutrino Experiment) according to which~\cite{Aker:2019uuj},
\be m_\beta < 1.1 ~\textrm{eV}. \ee In future, KATRIN is expected to probe values of $m_\beta$ as low as 0.2 eV. The Project 8 is another experiment which  is under construction and it  aims  to  achieve a  sensitivity of $m_\beta \sim 0.04$ eV~\cite{Esfahani:2017dmu}.

The Planck 2018 results put very stringent limit on the maximum value of the sum of active light neutrino masses as~\cite{Aghanim:2018eyx},
\be \Sigma = m_1 + m_2 + m_3 < 0.12~\textrm{eV}. \ee The value of $m_\beta$ in the MLRSM that we are considering will be the same as in the standard three neutrino picture. This is because, we have considered the mixing between the $W_L$ and $W_R$ gauge bosons to be negligible. In Fig.\ref{fig3}, we have plotted the correlation of $m_\beta$ against $\Sigma$. The left and the right panels are for NH and IH respectively. The yellow region is disfavored by Planck-2018 and the black solid and dotted lines correspond to the future sensitivity of KATRIN and Project 8 respectively. As already pointed out, these plots are independent of any other model parameters other than the the UPMNS mixing matrix and light neutrino masses.
From this figure, we can see that the region of parameter space that KATRIN is going to probe is in strong tension with cosmology. On the other hand, the Project 8 experiment will have the potential to probe the remaining IH region that is allowed by cosmology, thus enabling it to distinguish between NH and IH. Any measurement of $m_\beta$ outside the range shown in these figures will definitely be a signal of some other new physics.

In Fig.\ref{fig3b}, we have plotted the lightest neutrino mass (Red), $m_\beta$ (blue) and $\Sigma$ (Magenta) against the lightest neutrino mass. The left panel is for NH and the right panel is for IH. From this figure, we can see that the values of the lightest neutrino mass and $m_\beta$  become almost same for $m_1 \ge 0.02$ eV for NH and $m_3 \ge 0.1$ eV for IH. We can also see that there exists a predicted minimum value for both $m_\beta$ as well as $\Sigma$ from Fig.\ref{fig3b}. For NH, $m_{\beta~min} \sim $ 0.009 eV, $\Sigma_{min} \sim $ 0.06 eV and for IH, $m_{\beta~min} \sim $ 0.05 eV and $\Sigma_{min} \sim $ 0.1 eV.

It is also very instructive to plot the allowed intervals of $m_{ee}$ as a function of $ m_\beta $ and $\Sigma $. In Fig.\ref{fig4}, we have plotted these for type-I dominance with $r=0.01$ ($M_{\Delta_R^{++}} \sim 70 $ TeV  since $M_N$ = 500 GeV). The left and the right panels are for NH and IH respectively. The yellow region in the upper panel and the region shaded with gray dashes in the lower panel are disfavored by Planck-2018 and the region above the brown band is disfavored by the combined constraints from KamLAND-Zen and GERDA experiments. The horizontal black dotted and solid lines correspond to the future sensitivity of nEXO and LEGEND-1k respectively. The purple region corresponds to the future sensitivity of KATRIN and the vertical dashed black line corresponds to the future sensitivity of Project 8~\cite{Esfahani:2017dmu}. The blue dashed lines give the predictions for SM with three Majorana neutrinos. The orange region are the predictions for $m_{ee}$ in MLRSM with type-I dominance after including the LFV constraints. We have seen in the previous section that the LFV bounds do not put extra constraints for $r=0.01$. From this figure, the values of $m_\beta$ and $\Sigma$ in MLRSM corresponding to the cancellation region for NH (left panel) are around $\sim 0.01$ eV and $\sim 0.06$ eV respectively and this is the same as the standard three generation case. The major difference in the predictions for MLRSM with that of the three generation case is the presence of an enhancement in $m_{ee}$ values for $m_\beta \sim 0.01$ eV and $\Sigma \sim 0.06$ eV. This new region is allowed even after imposing the cosmological constraints and can be probed in LEEGEND-1k and nEXO. There is an enhancement in the predicted values of $m_{ee}$ in the case of IH as well (right panel) for $m_\beta \sim 0.07$ eV and $\Sigma \sim 0.1$ eV, though here, most of the enhancement region is disfavored by the current $0\nu\beta\beta$ bounds. Also, for IH, only a very narrow region with $m_\beta \sim 0.044-0.052$ eV is allowed after imposing the constraints from cosmology. The earlier mentioned tension between the region KATRIN is going to probe and the constraints from cosmology can be seen here as well. This is expected since the expressions for $m_\beta$ and $\Sigma$ are the same in MLRSM as well as the standard three neutrino picture.  All the allowed parameter space for IH can be probed completely by LEGEND-1k and Project 8 experiments.

In Fig.\ref{fig5}, we have shown similar plots for type-I dominance with $r=0.5$ ($M_{\Delta_R^{++}} \sim 1.4 $ TeV). Here also, the green regions show the predictions without including the LFV constraints whereas the orange region is after including the LFV constraints. As seen earlier, the contribution of the triplet scalar to LFV becomes important in this case and a major portion of the predicted values of $m_{ee}$ are disallowed. In particular, the cancellation region as well as the enhancement region for NH are disfavored by LFV constraints and a very narrow region remains allowed, as can be seen from the left panel of Fig.\ref{fig5}. In the case of IH (right panel of Fig.\ref{fig5}), the enhancement region is allowed by LFV, but most of it is disfavored by Kam-LAND-Zen + GERDA as in the previous figure. Also, since the regions that are disfavored by LFV in IH are disfavored by cosmology as well, the net allowed parameter space for type-I IH is the same for $r=0.01$ and $r=0.5$. In this case, the small allowed parameter region (orange region) for NH can be probed by nEXO whereas that for IH can be probed by LEGEND-1k and Project 8.

Figs.\ref{fig6} and \ref{fig7} show the correlations of $m_{ee}$ against $\Sigma$ and $m_\beta$ for $r=0.01$ and $r=0.5$ respectively in the case of type-II dominance. Here, we can see that there is no region with complete cancellation for $m_{ee}$ as we have already seen in the previous section. Also, there is considerable enhancement in the predicted values of $m_{ee}$ for NH, part of which is disfavored by $0\nu\beta\beta$ constraints. In addition, here also, the regions that are disfavored by LFV are disfavored by cosmology as well, making the allowed parameter space for type-II NH the same for $r=0.01$ and $r=0.5$. For IH, the predictions overlap almost completely with that for the standard three neutrino picture. Thus, for $r=0.01$ (Fig.\ref{fig6}), since there are no additional constraints from LFV, the allowed parameter space in the $m_\beta -m _{ee}$ plane is almost the same for IH in type-II LR as well as three neutrino picture. On the other hand, for $r=0.5$ (Fig.\ref{fig7}), the LFV constraints are severe in MLRSM and only a very small region around the point ($m_\beta \sim 0.045$, $m_{ee} \sim 0.047$) is allowed. For $r=0.01$, even though  most of the $m_{ee}$ parameter space for NH can be probed by nEXO, there is still a small region which is beyond the reach of nEXO. For IH with $r=0.01$ (Fig.\ref{fig6}) as well as $r=0.5$ (Fig.\ref{fig7}), all the allowed parameter space for IH is well within the reach of LEGEND-1k and Project 8.

\section{Constraining New Physics Parameters From LFV and Neutrino Mass Observables}

In the previous section, we discussed how the constraints from LFV affect the predictions for correlations among the absolute neutrino mass observables. In this section, we will see how the LFV constraints as well as the bounds on the values of the three neutrino mass observables restrict the allowed values of new physics parameters. Specifically, we will study the bounds on the heavy neutrino mass $M_N$, the parameter $r$ (or equivalently, the mass of the scalar $M_{\Delta_R}^{++}$) and the mass of the $SU(2)_R$ $W$ gauge boson, $M_{W_R}$. This also enables us to see the impact of varying $r$ and $M_N$ on the absolute neutrino mass observables. For all the plots in this section, oscillation parameters are varied in the $3\sigma$ ranges, Majorana phases are varied in the range $0-\pi$,$m_{lightest}$ is varied in the range $10^{-4}-0.1$ eV and $r$ is varied in the range $0-2$. Also, the upper and the lower panels are for type-I and type-II seesaw dominances, and the left and the right panels are for NH and IH, respectively in all the figures following.

  \begin{figure}
\begin{tabular}{cc}
\includegraphics[width=0.5\textwidth]{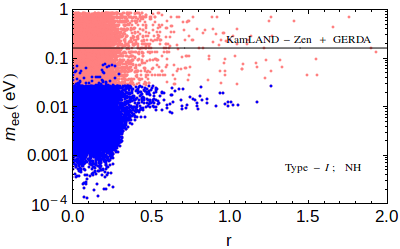}
\includegraphics[width=0.5\textwidth]{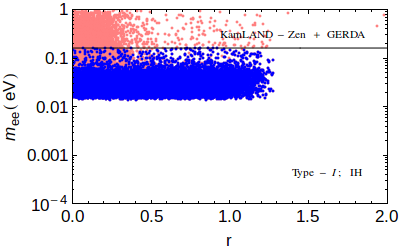}\\
\includegraphics[width=0.5\textwidth]{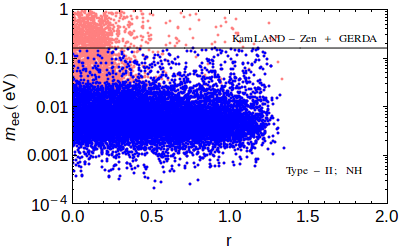}
\includegraphics[width=0.5\textwidth]{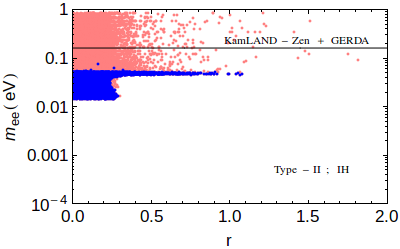}
\end{tabular}
\caption{The correlation of effective Majorana mass $m_{ee}$ against $r$. The upper (lower) panel is for type-I (type-II) dominance and the left (right) panel is for NH(IH). We have fixed $M_{W_R} = 7$ TeV. The pink points satisfy the LFV constraints whereas the blue points satisfy LFV constraints as well as the current bounds on $m_{ee}$, $m_\beta$ and $\Sigma$. Oscillation parameters are varied in the $3\sigma$ ranges, Majorana phases are varied in the range $0-\pi$, $M_N$ is varied in the range $ 0.1-30 $ TeV and $m_{lightest}$ is varied in the range $10^{-4}-0.1$ eV. The horizontal black line indicates the bound on $m_{ee}$ ($m_{ee} < 0.161$ eV).}\label{fig9}
\end{figure}

In Fig.\ref{fig9}, we have plotted the correlation of the effective Majorana mass $m_{ee}$ against $r$. Here, $M_N$ is varied in the range $ 0.1-30 $ TeV. The pink points satisfy the LFV constraints whereas the blue points satisfy LFV constraints as well as the current bounds on $m_{ee}$, $m_\beta$ and $\Sigma$. The bound on $m_{ee}$ comes from KamLAND-Zen and GERDA and is taken as $m_{ee} < 0.161$ eV (horizontal black line), which is the most liberal upper value corresponding to the lowest value of nuclear matrix elements. From this figure, we can see that there is an upper bound on $r$ (or correspondingly, a lower bound on $M_{\Delta_R^{++}}$) due to the constraints that we have imposed. These bounds for all the four cases are given in Table~\ref{table1}. Also, a very important result which is of special mention is that the bounds on $M_{\Delta_R^{++}}$ given in table-\ref{table1} is more stringent than the collider bounds (  $M_{\Delta_R^{++}}  > 660-760$ GeV)~\cite{Aaboud:2017qph} in the regime $M_N > 760$ GeV for all the four cases.

 We have seen earlier that for small $r$, LFV constraints are not that important and the parameter space allowed by $0\nu\beta\beta$ is allowed by LFV. In fact, from Fig.\ref{fig9}, we can see that there are some parameters for small r which are allowed by LFV but that gives 0nubb to be more than the current upper bound. In the case of type-I NH, cancellation region as well as predictions for $m_{ee} \sim 0.1$ eV are present only for the lower values of $r$. Specifically, for $r>0.26$, ${m_{ee}}_{max} \sim 0.03$ eV and for $r > 0.45$, ${m_{ee}}_{min} \sim 0.006$ eV. The pink points below the black line which are allowed by both LFV as well as $0\nu\beta\beta$ bounds are in fact disfavored by the bound on $\Sigma$ since those large values of $m_{ee}$ correspond to the quasi-degenerate region with $\Sigma > 0.12$ eV. For type-I IH, all the allowed values of $r$ can give predictions for $m_{ee}$ in the range $\sim 0.015 - 0.161$ eV. In type-II dominant case with NH, the predictions for $m_{ee}$ can take values in the range $ 2\times 10^{-4} - 0.161$ eV, depending upon the values of other parameters. Note that in Fig.\ref{fig2}, we saw that the minimum predicted value for $m_{ee}$ when $M_N =500$ GeV for type-II NH is $\sim 2 \times 10^{-3}$ whereas in Fig.\ref{fig9},  we can see that the predictions for $m_{ee}$ in type-II NH can be as low as $2 \times 10^{-4}$. This is because we have varied $M_N$ in the range $0.1-30$ TeV. From Eqn.~\ref{mee}, the contribution to $m_{ee}$ due to heavy neutrino exchange is inversely proportional to $M_N$ and hence, for large $M_N$, the values of $m_{ee}$ can be smaller. In type-II dominance with IH, for $r > 0.3$, the predicted values of $m_{ee}$ are around $\sim 0.05$ eV whereas for $r< 0.3$, $m_{ee}$ lies in the range $\sim 0.013 - 0.06$ eV. Thus, the maximum predicted values of $m_{ee}$ that are allowed by LFV and bounds on absolute neutrino mass observables lie considerably below the upper bound on $m_{ee}$ that we have taken. This is because the values of $m_{ee}$ in the range 0.06-0.161 eV (pink points below the black line) are disfavored by the Planck data since these high values of $m_{ee}$ come from large values of $m_{lightest}$ giving $\Sigma>0.12$ eV. Thus, from Fig.\ref{fig9}, we see that MLRSM in different limits gives slightly different upper limits on $r$ and lower limits on $M_{\Delta_R^{++}}$ and how all the three bounds (bounds from LVF, $0\nu\beta\beta$ and cosmology) conspire to give this upper limit, since had it been only LFV, much larger values of $r$ would have been allowed as shown by the pink points.

\begin{table}[ht]
 $$
 \begin{array}{|c|c|c|}
 
 \hline 
 
 \textrm{Model}& \textrm{Bound on } r & \textrm{Bound on } M_{\Delta_R}^{++} ( M_{\Delta_R^{++}} = \sqrt{2} M_N/r)   \\
 
 \hline
 
 Type-I NH & r < 1.26  & M_{\Delta_R^{++}} > 1.12 ~ M_N\\
 
 \hline
 
 Type-I IH & r < 1.28  & M_{\Delta_R^{++}} > 1.10 ~ M_N\\
 
 \hline
 
 Type-II NH & r < 1.35  & M_{\Delta_R^{++}} >1.05 ~ M_N \\
 
 \hline
 
 Type-II IH & r < 1.08  & M_{\Delta_R^{++}} > 1.31 ~ M_N \\

 \hline
 
 \end{array}
 $$
\caption{\small{ The upper bounds on $r$ and $M_{\Delta_R}^{++}$ (in terms of $M_N$) for the four different cases that we have studied.}}\label{table1}\end{table}


  \begin{figure}
\begin{tabular}{cc}
\includegraphics[width=0.5\textwidth]{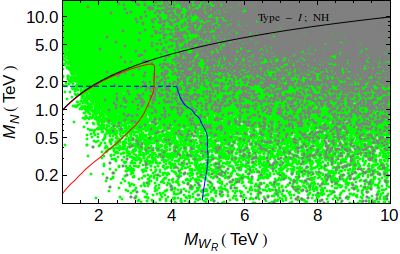}
\includegraphics[width=0.5\textwidth]{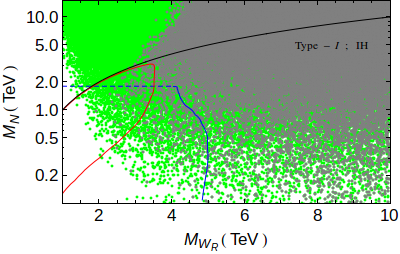}\\
\includegraphics[width=0.5\textwidth]{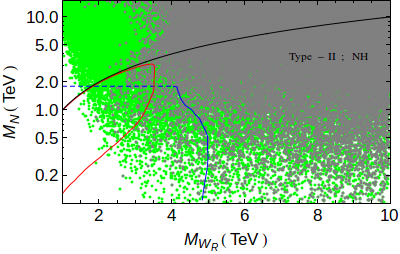}
\includegraphics[width=0.5\textwidth]{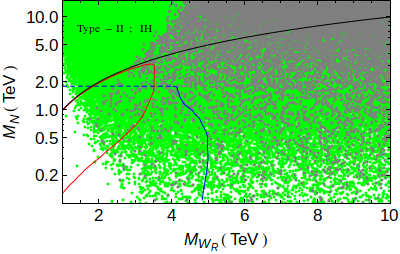}
\end{tabular}
\caption{Parameter space allowed by various constraints in the $M_{W_R}-M_N$ plane. $r$, $m_{lightest}$ and oscillation parameters are varied in the same ranges as in the previous plots. The green points satisfy the bound from $0\nu\beta\beta$ and the gray points satisfy the bounds on $0\nu\beta\beta$, $m_\beta$ and $\Sigma$ as well as the constraints from LFV. The region to the left of the blue and red lines are disfavored by the ATLAS and CMS analysis respectively. The black line corresponds to $M_N = M_{W_R}$. The upper (lower) panel is for type-I (type-II) dominance and the left (right) panel is for NH (IH). }\label{figmnmw}
\end{figure}

In Fig.\ref{figmnmw}, we have plotted the allowed parameter space in the $M_{W_R} - M_N$ plane. The oscillation parameters $r$ and $m_{lightest}$  are varied in the same ranges as in the previous plot. Here, the upper (lower) panel is for type-I (type-II) dominance and the left (right) panel is for NH (IH). The green points satisfy the bound from $0\nu\beta\beta$ whereas the gray points satisfy the bounds on $0\nu\beta\beta$, $m_\beta$ and $\Sigma$ as well as the constraints from LFV. Thus, only the gray points are allowed by all the constraints. The black line corresponds to $M_N = M_{W_R}$ and the collider experiments have given some bounds only below this line ($M_N < M_{W_R}$), where the signal with  two leptons and two jets ($W_R \rightarrow N_R ~l \rightarrow l~l~j~j $) in the final state is kinematically observable~\cite{Nemevsek:2011hz}. The region to the left of the blue line is disfavored by the ATLAS~\cite{Aaboud:2019wfg}  analysis done with the mass of $N_R$ in the range $0.1-18$ TeV. The region to the left of the red line is disfavored  by the  CMS~\cite{Sirunyan:2018vhk} analysis.

Here, we can see that both $0\nu\beta\beta$ and LFV can give constraints on the $M_{W_R}-M_N$ plane that are complimentary to the bounds from the collider. The region disfavored by $0\nu\beta\beta$ are shown by the white region in Fig.\ref{figmnmw}, which corresponds to the low values of $M_N$ and $M_{W_R}$. In reference~\cite{Dev:2013vxa}, the authors had studied the constraints from $0\nu\beta\beta$ in the $M_{W_R}-M_N$ plane for type-II dominance and they had observed that $0\nu\beta\beta$ provide constraints only for NH. Here, from Fig.\ref{figmnmw}, we can see that with the updated combined constraints on $m_{ee}$ from KamLAND-Zen and GERDA experiments \cite{Agostini:2018tnm}, $0\nu\beta\beta$ puts constraints on IH case as well. Also, it is to be noted that majority of the parameter space disfavored by $0\nu\beta\beta$ are also disfavored by the bounds from ATLAS and CMS experiments (Note that there is a small band corresponding to very small values of $M_{W_R}$, with $M_{W_R}<0.5$ TeV where large values of $M_N (\sim 10 \textrm{ TeV})$ are disfavored by $0\nu\beta\beta$, which is not shown in the figure). On the other hand, the combined constraints from LFV and $\Sigma$ disfavor certain regions (the green points lying above the red and blue lines) that are allowed by collider and $0\nu\beta\beta$. This is in the low $M_{W_R}$ and high $M_N$ region. It is interesting to note that the two cases with IH is relatively more constrained compared to the two NH cases. For instance, when $M_N = 10$ TeV, the lower bounds on $M_{W_R}$ are $\sim$ 2.4 TeV, 3.8 TeV, 2.8 TeV and 4 TeV respectively for type-I NH, type-I IH, type-II NH and type-II IH respectively. Thus LFV along with the absolute neutrino mass observables provide strong constraints on the values of $M_{W_R}$ in the regions that are not covered by the LHC analyses. 

\section{Conclusion}

In this work we have studied the correlations among the three absolute neutrino mass observables, $m_{ee}$, $m_{\beta}$ and $\Sigma$, in the context of minimal left-right symmetric model(MLRSM). Two phenomenologically interesting cases of type-I seesaw dominance as well as type-II seesaw dominance were considered for both NH as well as IH. From the study of correlations of absolute neutrino mass observables, we observed that the predictions for MLRSM can be very different compared to the standard three generation case. This is due to the contributions of the right handed current and the doubly charged scalar to $0\nu\beta\beta$. In particular, we found that there are new regions with considerable enhancement in the predicted values for $m_{ee}$ in all the cases except for the type-II dominance case with IH. Some of these newly allowed regions are already disfavored by the current bounds on $0\nu\beta\beta$ and a large part of the parameter space can be probed by the upcoming $0\nu\beta\beta$ experiments like LEGEND-1k and nEXO. Also, there exists a tension in the region of $m_\beta$ that the experiment KATRIN is going to probe and the bounds from cosmology. The planned experiment Project 8 with a sensitivity of $m_\beta \sim 0.04$ eV has the potential to probe the remaining IH region that is allowed by cosmology.

We have also taken into account the independent constraints coming from  charged lepton flavor violation and studied how these bounds are translated into the planes of absolute neutrino mass observables. We found that the constraints from LFV put very stringent limits on the prediction for absolute neutrino mass observables, especially for low mass range of the triplet scalar. We studied the effect of LFV for two different values of $M_{\Delta_R^{++}}$, viz, 70 TeV and 1.4 TeV. For $M_{\Delta_R^{++}} \sim 70$ TeV, LFV constraints had no effect on the predictions for absolute neutrino mass observables whereas for  $M_{\Delta_R^{++}} \sim 1.4 $ TeV, major part of the allowed parameter space was disfavored by LFV bounds. The mutually independent future measurements or null results from cosmology, single beta decay and $0\nu\beta\beta$ decay can throw more light into the underlying model.

We have also studied the constraints on the masses of the heavy particles using the bounds from  LFV and the three neutrino mass observables. In particular, we found that different models give different lower limit on $M_{\Delta_R}^{++}$ for a given value of $M_N$. As a general result, we found $M_{\Delta_R}^{++} > M_N$ (Table \ref{table1}) and this bound is more stringent than the collider bounds on $M_{\Delta_R^{++}}$  ($M_{\Delta_R^{++}} > 660-760$ GeV) in the regime $M_N > 760$ GeV. We have also determined the allowed parameter space in the $M_{W_R}-M_N$ planes and observed that LFV along with the absolute neutrino mass observables provide strong constraints on the values of $M_{W_R}$ in the regions that are not covered by the LHC analyses. The IH cases were found to be more constrained compared to the corresponding NH cases. All these can compliment the future searches for the heavy particles at the colliders. In addition, these studies can also be complimented with the theoretical constraints on the model parameter space considered, for instance, the constraints coming from absolute stability of the electroweak vacuum, tree level unitarity and perturbativity.

\section*{Acknowledgement}
The authors would like to thank Dr. Gulab Bambhania for his involvement during the initial stages of the work and Dr. Namit Mahajan for useful discussions.
  
\bibliographystyle{utphys}
\bibliography{tevportalnew}

\end{document}